\documentclass[aps,preprint, notitlepage, longbibliography]{revtex4-1}
\usepackage[dvips]{graphics,graphicx}
\usepackage{amsmath}
\usepackage{bm}

\begin{document}
\title{Decay of symmetry protected quantum states}
\author{Anna A. Bychek$^{1}$}
\author{Dmitrii N. Maksimov$^{1,2}$}
\author{Andrey R. Kolovsky$^{1,2}$}
\affiliation{
$^1$LV Kirensky Institute of Physics, Federal Research Center KSC SB RAS, 660036, Krasnoyarsk, Russia\\
$^2$ Siberian Federal University, Krasnoyarsk, 660041, Russia}
\date{\today}
\begin{abstract}
We study the decay of bosonic many-body states in the three well Bose-Hubbard chain where bosons in the central well can
escape into a reservoir. For vanishing inter-particle interaction this system supports a non-decaying
many-body state which is the antisymmetric
Bose-Einstein condensate with particles occupying  only the edge wells. In the classical approach this quantum state
corresponds to a symmetry protected non-decaying state which is stable even at finite interaction below a certain intensity threshold.
 Here we demonstrate that despite the classical counterpart is stable the antisymmetric Bose-Einstein
 condensate is always metastable at finite interatomic interactions due to quantum fluctuations.
\end{abstract}
\maketitle

\section{Introduction}

Dissipative quantum systems are of a large importance as
they pave a way for manipulating quantum matter for preparation of pure  \cite{diehl2008quantum} as well as highly entangled
\cite{lin2013dissipative} states, and implementation of quantum computations \cite{verstraete2009quantum}.
One particular example of dissipative quantum systems realized with cold atoms are open systems which can
exchange particles with a reservoir  \cite{witthaut2008dissipation,prosen2010exact,barmettler2011controllable,brantut2012conduction,barontini2013controlling,
krinner2015observation,lebrat2018band},
so, neither the energy nor the particle number is preserved. When an open system is coupled with two reservoirs with different chemical potentials,
it realizes an atomtronic analogue \cite{Pepino10} of semiconductor devices \cite{brantut2012conduction,ivanov2013bosonic,krinner2015observation,lebrat2018band,kolovsky2018landauer}.
Alternatively a quantum lattice system can be coupled with the environment only in a single lattice site as experimentally demonstrated in \cite{labouvie2016bistability}.

Here we consider
 decay of Bose particles from a triple quantum well with the central well coupled to environment,
 i.e. the bosonic particles are drained into a reservoir \cite{Kordas15}.
Despite its simplicity the three site-model BH
does not allow for exact analytic solution \cite{nemoto00, franzosi01, franzosi03, Kordas15,Bychek19a}.
In this paper we employ pseudoclassical approach \cite{mahmud2005quantum,
mossmann2006semiclassical,graefe2007semiclassical,
trimborn2008exact, kolovsky2009bloch,zibold2010classical,bychek2018noon,Bychek19, Bychek20}
which allows us to cast the problem into a form of coupled driven nonlinear oscillators.
From the pure classical perspective this
 system supports a non-decaying solution with equal intensities but opposite phases on the edge sites. Such a solution has a zero
 amplitude at the central site, and, therefore, is not directly coupled to the reservoir.
 This kind of localized solution, existing in the
  system despite the loss channels is allowed, is known as a bound state in the continuum (BIC) \cite{Hsu16}.
  In particular, the BIC to be considered in the present work is analogous to that supported by a pair of side-defects coupled to
  photonic crystalline waveguides \cite{Bulgakov11, Bulgakov11a}.  From the quantum mechanical perspective the discussed BIC is
  the antisymmetric Bose-Einstein condensate (BEC) with particles occupying only the edge sites.  The central problem to be addressed in the work is the
  account of the inter-particle interaction. We shall examine the stability of non-linear BIC in the classical
  regime and investigate the link between the classical and quantum solutions. It shall be demonstrated that even a classically stable nonlinear BIC
  undergoes a slow rate decay due to quantum fluctuations. Thus, the antisymmetric BEC is always a metastable state.
We shall show that the quantum fluctuations can be accurately
described within the pseudoclassical framework by the stochastic force emerging in the
nonlinear coupled oscillator model.

\section{Master equation and pseudoclassical approach}

\begin{figure*}[t]
\includegraphics[width=0.7\textwidth,height=0.36\textwidth,trim={0.0cm 0.0cm 0.0cm 0.0cm},clip]{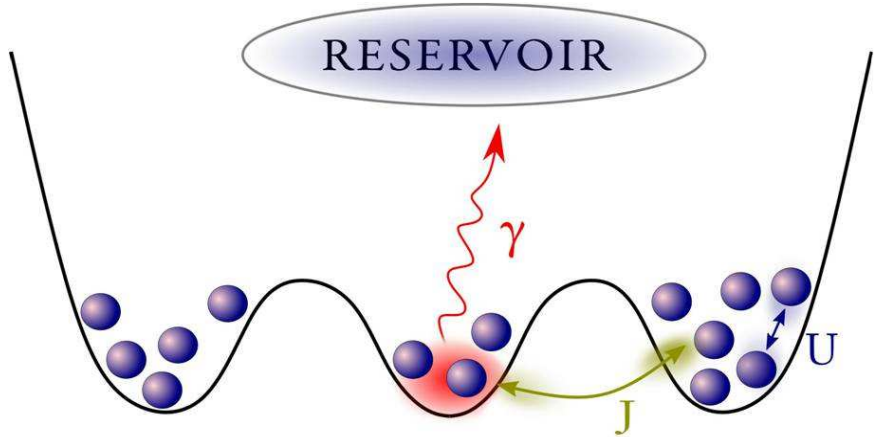}
\caption{Sketch of the system.} \label{fig1}
\end{figure*}

%
We consider a linear trimer of three coupled potential wells. The trimer is initially occupied by $N_0$ bosons which can tunnel between the wells as shown
in Fig. \ref{fig1}. The tunnelling  dynamics is controlled by the Bose-Hubbard Hamiltonian
\begin{align}\label{Hamiltonian}
&\widehat{{\cal H}}=
-\frac{J}{2}\sum_{\ell=1}^{2}\left(\hat{a}_{\ell+1}^{\dagger}\hat{a}_{\ell} +{\rm h.c.}\right) +\frac{U}{2}\sum_{\ell=1}^3\hat{n}_{\ell}
(\hat{n}_{\ell}-1),
\end{align}
where $\hat{a}_{\ell}^{\dagger}(\hat{a}_{\ell})$ is the creation (annihilation) operator at the $\ell_{\rm th}$ site, $\hat{n}_{\ell}$ is the number operator
at the $\ell_{\rm th}$ site,  $J$ is the interwell tunnelling rate and $U$ is the interaction constant.
By now the BH model has grown to one of the seminal model in physics of cold atoms which scopes quantum phase transitions \cite{greiner2002quantum},
the effects of Josephson oscillations \cite{Gati07, Esteve08}, atomic Bloch oscillations \cite{meinert2014interaction,fujiwara2019transport},
refill dynamics in the presence of induced losses, \cite{labouvie2015negative, labouvie2016bistability}, spontaneous breaking of the symmetry \cite{trenkwalder2016quantum},
and quantized current in the
engineered transport channel \cite{brantut2012conduction,krinner2015observation,lebrat2018band} to mention a few results relevant to
the present paper. Here, following \cite{Kordas15}, we assume that the central well is attached to an atom sink as shown in Fig. \ref{fig1}.
  Then the system dynamics is described by the density matrix $\widehat{{\cal R}}$ which obeys the master equation
\begin{equation}\label{master}
\frac{\partial \widehat{{\cal R}}}{\partial t}=-i[\widehat{{\cal H}},  \widehat{{\cal R}}]+
\widehat{{\cal L}}(\widehat{\cal R})
\end{equation}
with the loss operator of a Lindblad form
\begin{equation}\label{Liouvillian}
\widehat{{\cal L}}(\widehat{\cal R})=-\frac{\gamma}{2}
\left(\hat{a}_{2}^{\dagger}\hat{a}_{2}\widehat{\cal R }-2\hat{a}_{2}\widehat{\cal R }\hat{a}_{2}^{\dagger}
+\widehat{\cal R }\hat{a}_{2}^{\dagger}\hat{a}_{2} \right),
\end{equation}
where $\gamma$ is the loss rate.

The pseudoclassical approach is introduced by replacing each operator $\widehat{A}$ by its Weyl
symbol which is a function on the phase space \cite{McDonald88},
\begin{equation}
{\rm symb}[\widehat{A}]=A( a,{a}^*),
\end{equation}
with $a, a^*$ as the complex conjugated canonical variables
defined as the Weyl symbols of the annihilation and creation operators
\begin{equation}\label{conjugated}
{\rm symb}[\hat{a}]=a, \ {\rm symb}[\hat{a}^{\dagger}]=a^{*},
\end{equation}
where we omitted the subindex $\ell$ for simplicity. The Weyl symbols of an operator product of
two operators are computed via the Moyal star product of the Weyl symbols of the two operators
\begin{equation}\label{Moyal}
A\star
B=A\exp\left[\frac{\hbar}{2}\left(\frac{\partial^{\leftarrow}}{\partial
a}\frac{\partial^{\rightarrow}}{\partial a^*}-
\frac{\partial^{\leftarrow}}{\partial
a^*}\frac{\partial^{\rightarrow}}{\partial a} \right)\right]B.
\end{equation}
For instance, it is easy to see from Eq. (\ref{Moyal}) that the Weil symbol of the number operator is
\begin{equation}\label{number}
{\rm symb}[\hat{n}]=a^*\star a=|a|^2-\frac{1}{2}.
\end{equation}
The figure of merit in the pseudoclassical approach is the Weyl symbol of the density matrix known as the Wigner function
\begin{equation}
{\cal W}={\rm symb}[\widehat{{\cal R}}].
\end{equation}
Applying Eq. (\ref{Moyal}) to the master equation, Eq. (\ref{master}) one finds that the Wigner function obeys the following equation
\cite{Vogel88, Bortman95}
\begin{align}
& \frac{\partial {\cal W}}{\partial t}=-i\sum_{\ell=1}^{3}\left[U\left (1-|\alpha_{\ell}^2|\right)\left(\alpha_{\ell}\frac{\partial
{\cal W}}{\partial \alpha_{\ell}}
 -\alpha_{\ell}^*\frac{\partial
{\cal W}}{\partial \alpha_{\ell}^*}\right)
-\frac{U}{4}\left(\frac{\partial^3 \alpha_{\ell}^*{\cal W}}{\partial \alpha_{\ell}\partial{\alpha_{\ell}^*}^2}
-\frac{\partial^3 \alpha_{\ell}{\cal W}} {\partial \alpha_{\ell}^2\partial\alpha_{\ell}^*}
\right)
\right] \nonumber\\
& -i\frac{J}{2}\sum_{\ell=1}^{2}\left(
\alpha_{\ell+1} \frac{\partial {\cal W}}{\partial \alpha_{\ell}}
+\alpha_{\ell} \frac{\partial {\cal W}}{\partial \alpha_{\ell+1}}
-\alpha_{\ell+1}^* \frac{\partial {\cal W}}{\partial \alpha_{\ell}^*}
-\alpha_{\ell}^* \frac{\partial {\cal W}}{\partial \alpha_{\ell+1}^*}
\right)
 \nonumber \\
& +\frac{\gamma}{2}\left(\alpha_2\frac{\partial {\cal W}}{\partial \alpha_2} +2{\cal W} + \alpha_2^*\frac{\partial {\cal W}}{\partial \alpha_2^*}\right)
+\frac{\gamma}{2}\frac{\partial^2 {\cal W}}{\partial \alpha_2
\partial \alpha_2^*}.
\label{FP_W}
\end{align}
The above equation contains third order derivatives which do no allow to interpret it
as a Fokker-Plank equation with a positive definite or positive semidefinite diffusion matrix \cite{Vogel88}. 

The pseudoclassical limit of Eq. (\ref{FP_W}) is obtained by setting $N_0\rightarrow \infty$ while keeping $g=UN_0={\rm Const}$.
In what follows the quantity $g$ will be referred to as the macroscopic interaction constant. Let us apply the following substitution
\begin{equation}\label{substitution}
\alpha_{\ell}=\sqrt{N_0}a_{\ell}, \ \ \alpha^*_{\ell}=\sqrt{N_0}a^*_{\ell} .
\end{equation}
Then Eq. (\ref{FP_W}) transforms to
\begin{align}
& \frac{\partial {\cal W}}{\partial t}=-i\sum_{\ell=1}^{3}\left[g\left (\frac{1}{N_0}-|a_{\ell}^2|\right)\left(a_{\ell}\frac{\partial
{\cal W}}{\partial a_{\ell}} -a_{\ell}^*\frac{\partial \cal {\cal W}}{\partial a_{\ell}^*}\right) \right]
\nonumber\\
& -i\frac{J}{2}\sum_{\ell=1}^{2}\left(
a_{\ell+1} \frac{\partial  {\cal W}}{\partial a_{\ell}}
+a_{\ell} \frac{\partial {\cal W}}{\partial a_{\ell+1}}
-a_{\ell+1}^* \frac{\partial {\cal W}}{\partial a_{\ell}^*}
-a_{\ell}^* \frac{\partial {\cal W}}{\partial a_{\ell+1}^*} \right)
 \nonumber \\
& +\frac{\gamma}{2}\left(a_2\frac{\partial {\cal W}}{\partial a_2} + 2{\cal W}+a_2^*\frac{\partial {\cal W}}{\partial a_2^*}\right)
+\frac{\gamma}{2N_0}\frac{\partial^2 {\cal W}}{\partial a_2
\partial a_2^*}+{\cal O}(N_0^{-2}).
\label{FP_W_classical}
\end{align}
Neglecting ${\cal O}(N_0^{-2})$ term we arrive at a true Fokker-Plank equation where the first term in the third line  can be
viewed as dissipation while the second term in the same line is diffusion \cite{Bychek19, Bychek20}.

The dynamics under the Fokker-Planck equation, Eq. (\ref{FP_W_classical}) can be unravelled into a set of dissipative Langevin equations
\begin{align}
& id{a}_1=\left(-\frac{J}{2}a_2 +g|a_1|^2 a_1\right)dt, \nonumber \\
& id{a}_2=\left[-\frac{J}{2}\left(a_1+a_3\right) +g|a_2|^2 a_2 -i\frac{\gamma}{2} a_2\right]dt
+\sqrt{\frac{\gamma}{2N_0}}d\xi, \nonumber \\
& id{a}_3=\left(-\frac{J}{2}a_2 +g|a_3|^2 a_3\right)dt, \label{Langevin}
\end{align}
 where $d\xi$ is the complex white noise
 \begin{equation}\label{noise}
\overline{d\xi}=0, \ \overline{d\xi^{*} d\xi}=dt, \ \overline{d\xi d\xi}=0.
 \end{equation}
Notice that compared to Eq. (\ref{FP_W}) in Eq. (\ref{Langevin}) we omitted the "self-energy" term proportional to $g/N_0$. This can be done as the oscillating factor $\exp(-igt/N_0)$ can be absorbed into the noise Eq. (\ref{noise}) without changing its correlation properties.

Let us assume for a moment that there is no noise term in Eq. (\ref{Langevin}). Then Eq. (\ref{Langevin}) has a antisymmetric solution decoupled from the lossy site
\begin{equation}
{\bm a}_{\rm BIC}(t)=
e^{-igIt}
\left(
\begin{array}{c}
\sqrt{I} \\
0 \\
-\sqrt{I}
\end{array}
\right) \label{BIC}
\end{equation}
where intensity $I$ can be linked to the mean population of the edge sites $\overline{n}_{1,2}=IN_0+{1}/{2}$. By examination of
Eq.~(\ref{Langevin}) one immediately identifies the three factors affecting the decay dynamics of this state:

(i) The stability of the BIC. If Eq.(\ref{BIC}) is unstable, it can be destroyed by
small perturbations.

(ii) The initial condition for solving Eq. (\ref{Langevin}). In more detail, we expect that the decay rate is dependent
on how close the initial condition is to the symmetry protected BIC, Eq. (\ref{BIC}).
Moreover in establishing quantum to classical correspondence one can not deal with a single trajectory but with
the ensemble of trajectories whose initial conditions are determined by the initial many-body quantum state of the system \cite{kolovsky2009bloch}.

(iii) The noise term in Eq. (\ref{Langevin}) inversely proportional to $\sqrt{N_0}$. The
noise can perturb even an intrinsically stable state driving it out of equilibrium.
Notice
that even though the reservoir does not supply particles into the system a stochastic driving force is still present in Eq. (\ref{Langevin}).
Physically, this intrinsic noise is nothing but the quantum fluctuations arising from the noncommutativity
of creation and annihilation operators. The noise term is important for the correct  application of the pseudoclassical approach.
For example, in the paradigm problem of the decaying quantum oscillator it ensures that the oscillator does not decay
below its ground energy.

In the next section we discuss each of these factors in more detail.

\section{Decay of the antisymmetric state}
\subsection{Stability analysis}

First we analyze the stability of the solution  ${\bm a}_{\rm BIC}(t)$ for $g\ne0$.
Using the standard stability analysis \cite{lichtenberg2013regular}
the stability of
this solution can be examined by analyzing the matrix
\begin{equation}\label{M}
 \widehat{M}=
 \left(
 \begin{array}{cccccc}
 gI & -J/2 & 0 & gI & 0 & 0 \\
 -J/2 & -i\gamma/2-gI & -J/2 & 0 & 0 & 0 \\
 0 & -J/2 & gI  & 0 & 0 & gI \\
 -gI & 0 & 0 & gI & J/2 & 0 \\
 0 & 0 & 0 &J/2 & -i\gamma/2+gI & J/2 \\
 0 & 0 & -gI & 0 & J/2 & gI   \\
 \end{array}
\right).
\end{equation}
If the imaginary part of all eigenvalues of $ \widehat{M}$ is non-positive, the BIC solution is stable.
 Fig. 1 (a) shows the imaginary parts of the eigenvalues  as the function of $gI$ for $J=1$ and $\gamma=0.4$. It is seen that
 in this case the stability threshold corresponds to $gI=0.2$. Above the threshold any tiny imbalance in the population
 of the edge sites will leads to excitation of the symmetric modes and the BIC looses its intensity.
 This process is exponential in time resulting in a rapid drop
 of intensity at the initial stage. However, once the stability threshold is crossed the solution ${\bm a}_{\rm BIC}(t)$ stabilizes
 at a certain value of intensity $I_{\rm st}$,
\begin{equation}\label{thres}
I_{\rm st}=I_{\rm st}(\gamma,g) .
\end{equation}
In what follows we shall refer to Eq.~(\ref{thres}) as the stabilization level. We mention that, as expected, the stabilization level is
 approximately inversely proportional to $g$ yet it is always smaller than the stability threshold deduced from Eq.~(\ref{M}).
 The described scenario is exemplified by thin  solid lines in Fig.~2 (d, e) where, to provoke the symmetry breaking, we introduced a
 tiny population imbalance $\sim10^{-3}$ in the initial BIC state. The exponential decrease of intensity for $g$ above the stability threshold
 and the effect of stabilization is clearly seen in the figure.
\begin{figure*}[t]
\includegraphics[width=1\textwidth,height=0.72\textwidth,trim={3.0cm 0.0cm 0.0cm 0.0cm},clip]{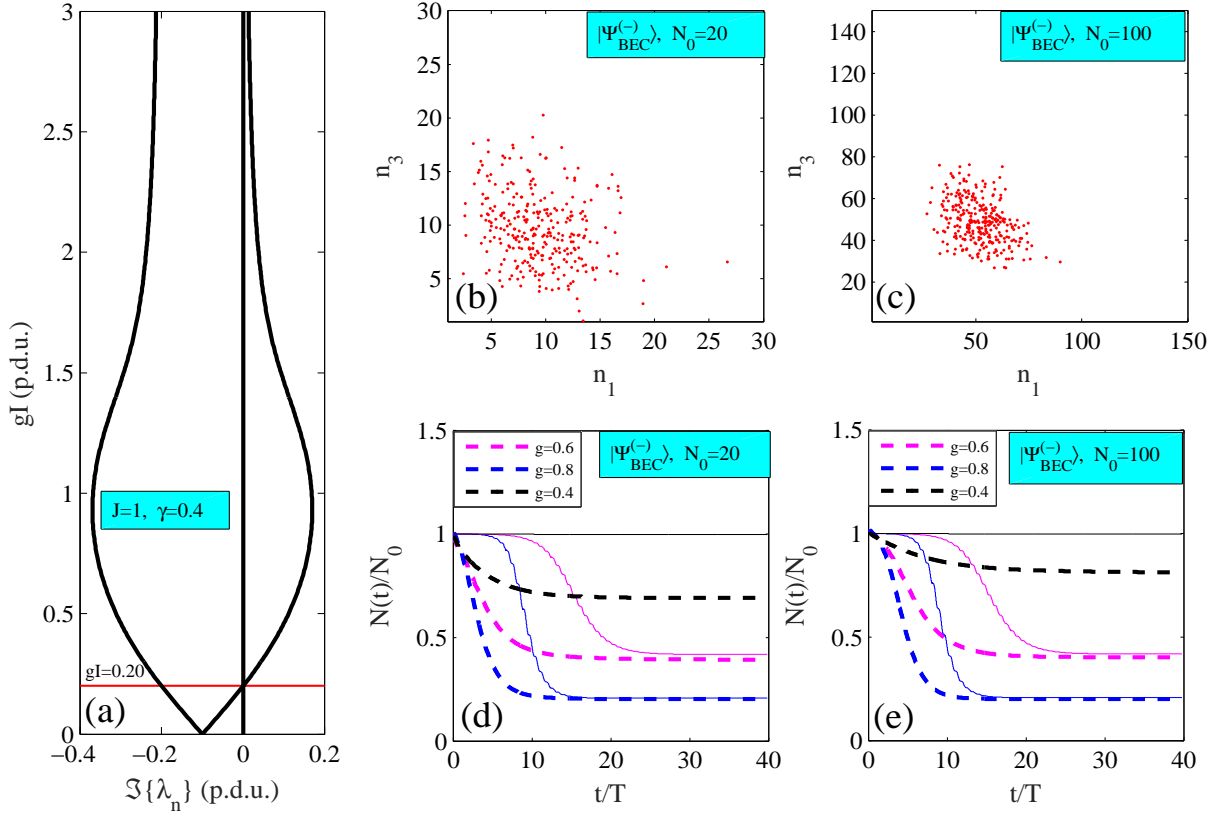}
\caption{(a) Imaginary parts $\Im\left\{\lambda_n \right\}$ of eigenvalues of matrix Eq. (\ref{M}) for $\gamma=0.4$.
(b, c) Initial conditions for the antisymmetric BEC in the space of populations of the first, $n_1$ and the third, $n_3$ sites for (b) $N_0=20$ and (c) $N_0=100$.
(d, e) Decay dynamics of the antisymmetric state with (d) $N_0=20$ and (e) $N_0=100$.
Thick dashed lines show the total number of particle against time for antisymmetric BEC initial conditions shown in subplot (d, c).
Thin solid lines show the decay of the classical
BIC state.} \label{fig2}
\end{figure*}

\subsection{Quantum ensemble}

Before simulating the decay of truly quantum states we have to introduce an ensemble classical initial condition
corresponding to a quantum state loaded into the system.
This, can be done by using the Husimi ${\cal Q}$-function,
\begin{equation}\label{Husimi}
{\cal Q}({\bm \alpha})=\frac{1}{\pi^3}\langle{\bm \alpha}| \widehat{\cal R}|{\bm \alpha}\rangle,
\end{equation}
where $|{\bm \alpha}\rangle$ is the Glauber coherent state,
\begin{equation}\label{coherent}
|{\bm \alpha}\rangle=e^{-\frac{|\alpha_1|^2+|\alpha_2|^2+
|\alpha_3|^2}{2}}e^{\alpha_1\hat{a}_1^{\dagger}+\alpha_2\hat{a}_2^{\dagger}+\alpha_3\hat{a}_3^{\dagger}}|{\rm vac} \rangle
\end{equation}
with ${\bm \alpha}=\left\{ \alpha_1, \alpha_2, \alpha_3\right\}$.  At first, for the initial state we choose an antisymmetric $N$-particle BEC.
However, for the future convenience below we
present a single formula for  both symmetric, $|\Psi^{(+)}_{\rm BEC}\rangle$ and antisymmetric $|\Psi^{(-)}_{\rm BEC}\rangle$
condensates
\begin{equation}\label{BEC}
|\Psi^{(\pm)}_{\rm BEC}\rangle=\frac{1}{\sqrt{2^NN!}}\left({\hat{a}_1^\dag \pm \hat{a}_3^\dag}\right)^N |{\rm vac} \rangle.
\end{equation}
After applying the Husimi transformation Eq. (\ref{Husimi}) one finds
\begin{equation}\label{BECH}
{\cal Q}^{(\pm)}_{\rm BEC}({\bm \alpha})=\frac{|\alpha_1\pm\alpha_3|^{2N}}{\pi^3 2^N(N!)}e^{-|\alpha_1|^2-|\alpha_2|^2-
|\alpha_3|^2}.
\end{equation}
By applying the acception-rejection method \cite{kolovsky2009bloch} we generate the ensembles of the initial conditions
 according to the distribution function (\ref{BECH}) for $N_0=20$ and $N_0=100$, see panels (b) and (c) in Fig.~2, and
 then simulate the system dynamics. The result is depicted by the thick dashed lines in panels (d) and (e). Remarkably,
  in the unstable cases ($g>0.4$) we get the same fraction of bosons which is left in the system as it is predicted by
  the pure classical stability analysis given in the previous subsection. In the stable case $g=0.4$, however, we observe essential
  deviations. These can be understood by  noticing that every initial condition ${\bf a}(t=0)$ from the quantum ensemble is a
  superposition of the system linear eigenmodes,
 \begin{equation}
{\bm b}_{1}=
\left(
\begin{array}{c}
\frac{1}{\sqrt{2}} \\
0 \\
\frac{-1}{\sqrt{2}}
\end{array}
\right), \
{\bm b}_{2}=
\left(
\begin{array}{c}
\frac{1}{{2}} \\
\frac{1}{\sqrt{2}} \\
\frac{1}{{2}}
\end{array}
\right), \
{\bm b}_{3}=
\left(
\begin{array}{c}
\frac{1}{{2}} \\
\frac{-1}{\sqrt{2}} \\
\frac{1}{{2}}
\end{array}
\right) \label{eigenmodes}
\end{equation}
where the first eigenmode  obviously corresponds to the symmetry protected BIC while the other two modes are coupled to the reservoir and decay within the characteristic time $2\pi/\gamma$.  Thus, the solid and dashed lines in Fig.~2(d-e) may coincide only in the limit $N_0\rightarrow\infty$ where the quantum ensemble shrinks to the single point.

\subsection{The Role of the Noise}

\begin{figure*}[t]
\includegraphics[width=0.8\textwidth,height=0.48\textwidth,trim={0.0cm 0.0cm 0.0cm 1.4cm},clip]{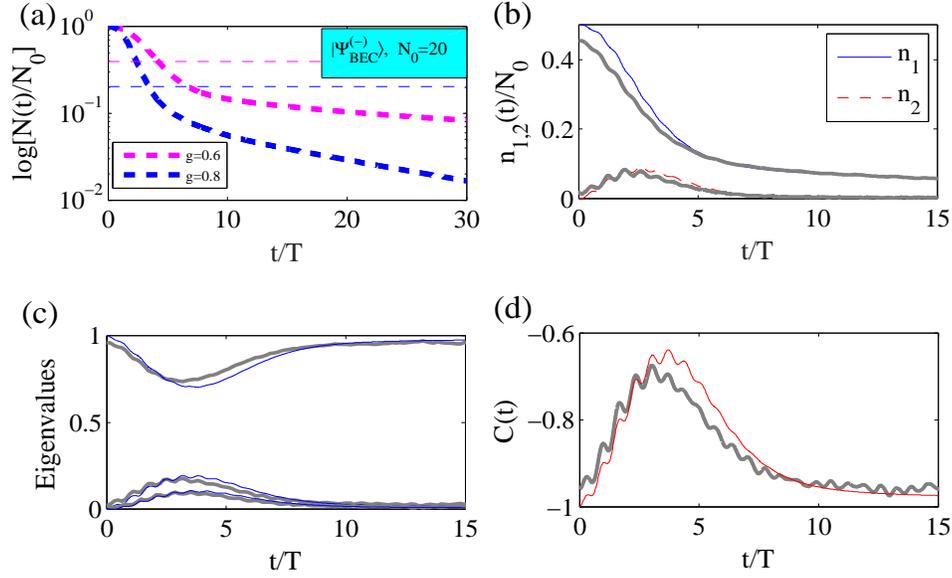}
\caption{Decay of the antisymmetric condensate for $N_0=20$.
(a) Logarithmic plot of the full population against time obtained by solving the master equation Eq.
(\ref{master}). The thin dash lines show the stabilization levels from Fig. \ref{fig2} (d).
(b) Populations of the first and the second sites against time. Here and in the panels (c, d) thin lines are quantum simulations while thick grey lines are
the results computed with the  pseudoclassical approach, $g=0.8$.
(c) Normalized eigenvalues of the single particle density matrix Eq. (\ref{single_density}), $g=0.8$.
(c) Correlation function Eq. (\ref{corfun}), $g=0.8$. } \label{fig3}
\end{figure*}

Now let us return to the Langevin dynamics governed by Eq. (\ref{Langevin}) where one could expect that even a stable
 antysimmetric BIC state Eq. (\ref{BIC}) is subject to decay. To test this conjecture we solve numerically both the Langeven equation and
  the exact master equation, Eq. (\ref{master}). The results are shown in Fig.~\ref{fig3}. In Fig. \ref{fig3}(a) we depict the exact quantum
   solution for the total population. One can see that unlike the classical solutions in Fig. \ref{fig2} the populations now continues to
   decay even after crossing the stabilization level. This decay is still exponential, however, with much smaller rate. In Fig. \ref{fig3}(b)
    we compare  the population dynamics for the first and the second sites obtained by the pseudoclassical and quantum approaches. One can see that
    the two results are in a good agreement. This supports our conjecture that the noise destroys the classical symmetry protected BIC.

To look at the decay dynamics in more detail we compute the single particle density matrix $\hat{\rho}$, whose matrix elements are defined as
\begin{equation}\label{single_density}
\rho_{\ell,\ell'}={\rm Tr}\left(\hat{a}_{\ell}^{\dagger}\hat{a}_{\ell'}\widehat{\cal R }\right).
\end{equation}
In the pseudoclassical framework this matrix corresponds to the correlation functions
\begin{equation}
\rho_{\ell,\ell'}=\langle a_{\ell}^*a_{\ell'}\rangle-\frac{1}{2}\delta_{\ell,\ell'},
\end{equation}
where the pointy brackets designate the ensemble average over Langevin trajectories. The single particle density matrix allows us to test whether the quantum state remains a BEC during the decay \cite{Mueller06}. Namely, if all but one eigenvalues are zero the system
is a condensate state. Another related quantity is the correlation function \cite{Kordas15},
\begin{equation}\label{corfun}
C(t)=\frac{\rho_{1,3}}{\sqrt{ \rho_{1,1}\rho_{3,3}}}
=\frac{\langle a_1^* a_3 \rangle}{\sqrt{\langle |a_1|^2 \rangle \langle |a_3|^2 \rangle}}.
\end{equation}
If $C(t)=-1$ the system is an antisymmetric condensate and if $C(t)=1$, then the condensate is symmetric. In Fig. \ref{fig3} (c) we show the
normalized eigenvalues of $\hat{\rho}$, while in Fig. \ref{fig3} (d) we plotted the correlation functions, Eq. (\ref{corfun}). Both subplots are consistent with the decay dynamics described above: First, the system is rapidly departs from antisymmetric BEC. Then,
after the stabilization level is crossed the system recoheres into antisymmetric BEC again and slowly decays as the metastable
antisymmetric solution.


\subsection{Symmetric BEC}

\begin{figure*}[t]
\includegraphics[width=0.8\textwidth,height=0.48\textwidth,trim={0.0cm 0.0cm 0.0cm 1.8cm},clip]{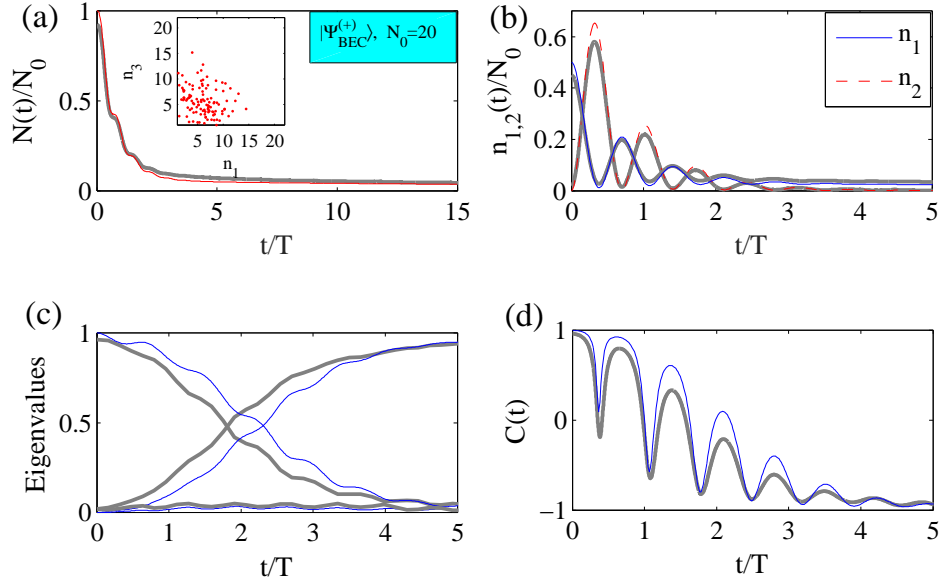}
\caption{Decay of symmetric condensate for $g=0.8$ and  $N_0=20$.
(a) The full population against time as obtained by solving the master equation Eq. (\ref{master}).
(b) Populations of the first and the second wells against time; thin line -- quantum simulations, thick grey lines -- pseudoclassical approach.
(c) Normalized eigenvalues of the single particle density matrix Eq. (\ref{single_density}).
(d) Correlation function Eq. (\ref{corfun}).} \label{fig4}
\end{figure*}

To see the full picture in this section we also present results on the decay of symmetric BEC state, Eq. (\ref{BEC}). One can see from Fig. \ref{fig4}
that in the course of evolution the symmetric BEC state rapidly drops intensity and decoheres into a fractional condensate
with two non-zero eigenvalues of the single particle density matrix. Eventually, only a tiny fraction of
the initial population survives after the system transits into a pure antisymmetric BEC well below the stability threshold.
Again we see a good coincidence between the quantum and pseudoclassical results.

\begin{figure*}[t]
\includegraphics[width=1\textwidth,height=0.8\textwidth,trim={0.0cm 0.0cm 0.0cm 0cm},clip]{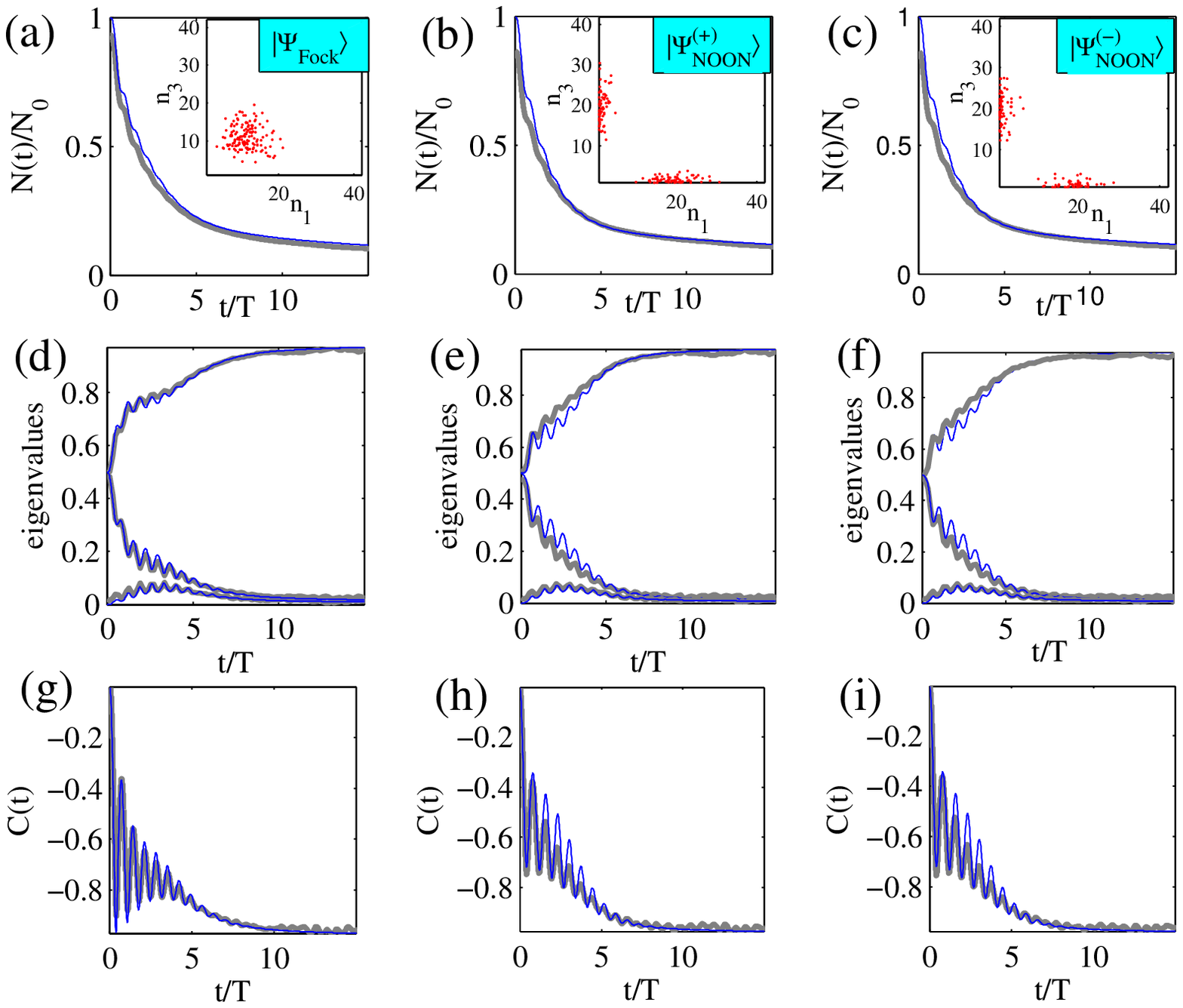}
\caption{Decay of fractional condensates, $N_0=20, \ T=2\pi \ {\rm (p.d.u.)}$; thick grey lines show the result by the pseudoclassical
approach, thin blue - lines the solutions of the master equations. (a-c) Dynamics of the full population. The insets show
the ensembles of initial conditions. (d-f) Normalized eigenvalues of the single particle density matrix for (d) Fock state, (e) symmetric NOON state, and (f) antisymmetric NOON state. (g-i) Correlation function, Eq. (\ref{corfun}) for (g) Fock state, (h) symmetric NOON state, and (i) antisymmetric NOON state.} \label{fig5}
\end{figure*}

\section{Decay of fractional condensates}
Next we examine the decay dynamics of fractional condensate states. The fractional condensate states are defined as those having more than one non-zero eigenvalues of the single particle density matrix \cite{Mueller06}. The most obvious example of a fractional condensate is a Fock state
\begin{equation}\label{Fock}
|\Psi_{\rm Fock}\rangle=\frac{1}{(N/2)!}\left({\hat{a}_1^\dag\hat{a}_3^\dag}\right)^{(N/2)} |{\rm vac} \rangle,
\end{equation}
where $N_0/2$ particles occupy the first site and the rest $N_0/2$ particle the third site. Directly applying Eq. (\ref{Husimi}) one finds
\begin{equation}\label{FockH}
{\cal Q}_{\rm Fock}({\bm \alpha})=\frac{|\alpha_1|^N|\alpha_3|^N}{\pi^3 [(N/2)!]^2}e^{-|\alpha_1|^2-|\alpha_2|^2-
|\alpha_3|^2}.
\end{equation}

Another, less trivial, example is the (anti-)symmetric NOON state which is a Shr\"odinger cat state of $N$ bosons in two wells,
\begin{equation}\label{NOON}
|\Psi^{(\pm)}_{\rm NOON}\rangle=\frac{1}{\sqrt{2(N!)}}\left[(\hat{a}_1^\dag)^N\pm(\hat{a}_3^\dag)^N\right] |{\rm vac} \rangle.
\end{equation}
This state has the following ${\cal Q}$-function
\begin{equation}\label{NOONH}
{\cal Q}^{(\pm)}_{\rm NOON}({\bm \alpha})=
\frac{|\alpha_1|^{2N}+|\alpha_3|^{2N}\pm(\alpha_1\alpha_3^*)^N\pm(\alpha_1^*\alpha_3)^N}{\pi^3 2(N!)}e^{-|\alpha_1|^2-|\alpha_2|^2-
|\alpha_3|^2}.
\end{equation}
In Fig. \ref{fig5} we show the simulation results by both pure quantum and pseudoclassical approaches.
 One can see that despite the profound difference between the Fock state and the (anti-)symmetric NOON states clearly
 seen in  insets in Fig. \ref{fig5} (a-c), the decay dynamics is essentially identical. In all cases we see a rapid decay
 of a fractional state below the stability threshold after which the system recoheres to the antisymmetric BEC having lost
  the major part of the initial population. As before we see a good accuracy of the pseudoclassical approach.

\section{Summary and Conclusion}

We have examined the decay dynamics of quantum states with a definite number of bosons in three well open Bose-Hubbard model.
It is demonstrate that the stability of the quantum state can be predicted from the classical perspective. The decay
scenarios are drastically different depending on whether the solution is stable in the pseudoclassical limit.
In particular it is shown that in the pseudoclassical regime the antisymmetric BEC state is mapped to
a symmetry protected bound state in the continuum (BIC). The BIC is only stable below a certain intensity threshold.
Above the classical stability threshold the antisymmetric BEC rapidly decays and decoheres due to inter-particle interactions.
Once the population has dropped below the threshold, however, the system recoheres to the antisymmetric BEC which decays at much slower rate due
to the quantum fluctuations. It is demonstrated that the quantum fluctuations can be accurately described
in the pseudoclassical framework by introducing a stochastic force with amplitude inversely proportional to the square root of the initial number
of particles.

The pseudoclassical approach has been applied to several types of initial states with initial population only at the edge sites.
Besides the antisymmetric BEC we have studied the decay of symmetric BEC, Fock, (anti-)symmetric NOON states. In all cases the
initial bosonic cloud rapidly losses population to the reservoir and the decay can only slow down well below the
classical stability threshold when the system recoheres to the metastable antisymmetric BEC. In all cases we observed
a good coincidence between the numerical data obtained by the pseudoclassical approach and direct quantum simulations.

Recently, we have seen a surge of interest to decay dynamics
of two-photon states \cite{Crespi15, Chen18, Zhang20, Poddubny20}. We believe that
the approach presented here provides the key to understanding the decay dynamics in the other
solvable limit, namely, pseudoclassical regime.
Finally, we would like to outline the future fork ensuing from the present paper.
It is remains a question whether the asymptotic law of below threshold decay can be derived from the Langevin
equations, Eq. (\ref{Langevin}) or the corresponding Fokker-Plank equation.
We speculate that this problem may pose an interesting topic for future research.

This work has been supported by Russian Science
Foundation through grant N19-12-00167. We appreciate discussions with A.F. Sadreev and E.N. Bulgakov.
We are also grateful to G.P. Fedorov for his critical reading of the manuscript.

\bibliography{Open_systems}

\end{document}